\documentclass[letterpaper,11pt]{article}
\usepackage{fullpage}
\usepackage[hmargin=1in,vmargin=1in]{geometry}
\usepackage[english]{babel}
\usepackage[utf8]{inputenc}
\usepackage[T1]{fontenc}
\usepackage{amsmath,amsthm}
\usepackage{algorithmic}
\usepackage{algorithm}
\usepackage{enumerate}  
\usepackage{subfig}
\usepackage{amssymb}
\usepackage{lipsum}
\usepackage{mathtools}
\usepackage{graphicx,color}
\usepackage{authblk}
\usepackage{hyperref}
\usepackage{booktabs} 
\usepackage[dvipsnames]{xcolor}
\usepackage{lineno}
\usepackage{graphicx}
\usepackage{mathrsfs,bm}
\usepackage{color}
\usepackage{float}

\newtheorem{lemma}{Lemma}
\newtheorem{theorem}{Theorem}
\newtheorem{definition}{Definition}
\newtheorem{proposition}{Proposition}
\newtheorem{claim}{Claim}
\newtheorem{corollary}{Corollary}

\newcommand{\nbd}[1]{\mbox{Nbd}(#1)}
\newcommand{\degree}[1]{\mbox{deg}(#1)}
\newcommand{\proby}[1]{P\left[ #1 \right]}

\newcommand{\probm}[1]{P_w[ #1 ]}
\newcommand{\pmd}{\mathcal{P}}
\newcommand{\probmt}[1]{\mathcal{P}^t[ #1 ]}

\newcommand{\thit}{t_{\mbox{\scriptsize hit}}}

\newcommand{\sink}{u_{\mbox{\scriptsize s}}}
\newcommand{\Var}{\mbox{Var}}
\newcommand{\expt}[1]{{\mbox{E}\left[#1\right]}}
\newcommand{\ind}[1]{{\mbox{\bf 1}_{\{#1\}}}}

\newcommand{\ignore}[1]{}

\def\RR{\mathbb{R}}
\def\nn0{\mathbb{N} \cup \{0\}}
\def\nonsinks{V \setminus \{\sink\}}

\providecommand{\keywords}[1]{\textbf{\textit{Keywords:}} #1}

\begin{document}

\title{A Stochastic Process on a Network with Connections to Laplacian Systems of Equations} 

\author[1]{Iqra Altaf Gillani}
\author[1]{Amitabha Bagchi}
\author[2]{Pooja Vyavahare}
\affil[1]{\{iqraaltaf,bagchi\}@cse.iitd.ac.in\\ Department of Computer Science and Engineering, IIT Delhi}
\affil[2]{poojav@iittp.ac.in\\ Department of Electrical Engineering, IIT Tirupati}

\date{}
\maketitle
\begin{abstract}
We study an open discrete-time queueing network that models the collection of data in a multi-hop sensor network. We assume data is generated at the sensor nodes as a discrete-time Bernoulli process. All nodes in the network maintain a queue and relay data, which is to be finally collected by a designated sink. We prove that the resulting multi-dimensional Markov chain representing the queue size of nodes has two behavior regimes depending on the value of the rate of data generation. In particular, we show that there is a non-trivial critical value of data rate below which the chain is ergodic and converges to a stationary distribution and above which it is non-ergodic, i.e., the queues at the nodes grow in an unbounded manner. We show that the rate of convergence to stationarity is geometric in the sub-critical regime. We also show the connections of this process to a class of Laplacian systems of equations whose solutions include the important problem of finding the effective resistance between two nodes, a subroutine that has been widely used to develop efficient algorithms for a number of computational problems. Hence our work provides the theoretical basis for a new class of distributed algorithms for these problems. 
\end{abstract}


\section{Introduction}
\label{sec:intro}
We study a stochastic process arising from a natural routing and scheduling scheme used to collect data from sensor nodes over multi-hop relay networks~\cite{Heinzelman-HICSS:2000,Kamra-SIGCOMM:2006,An-JNW:2015}. We model the sensor network as a graph, some of whose vertices produce data packets according to a discrete-time Bernoulli process. One node is designated as a sink that has to collect the data generated in the network and all other nodes relay the data. Each node maintains a queue and relays at most one packet in a time slot in the manner of the ``gossip'' models widely studied in the networking and distributed computing literature~\cite{Kempe-FOCS:2003}\cite{Boyd-INFOCOM:2005}\cite{Mosk-PODC:2006}. The packet is relayed to a random neighbor, in the manner of a random walk on the graph. Our model is, therefore, an open discrete-time queueing network whose interconnections are described by an undirected (simple) graph. 


Due to the relationship with the data collection task we call our process the {\em Data Collection Process} defined on a graph $G = (V,E)$ equipped with a positive edge-weight function $w : E \rightarrow \RR_+$. The process takes two parameters, a {\em relative rate vector} $\bm{J}\in (0,1)^{|V|}$ and a {\em rate} $\beta \in (0,1)$; we assume that node $v \in V$ produces a packet with probability $\beta \bm{J}(v)$ in a given time slot. For a given relative rate vector, the process has two behavior regimes and undergoes a sharp transition between these two regimes, the controlling parameter being the rate $\beta$. Specifically, we will show that for a critical value $\beta^*$ we have that when $\beta >\beta^*$ the process is non-ergodic, and the size of the queues grows to infinity, whereas, when $\beta<\beta^*$ process is ergodic such that all queues are almost surely finite and the system converges to a stationary distribution. For this latter regime, we also show that the rate of convergence is geometric, i.e., the Data Collection Process is geometrically ergodic. 
 
For $\beta<\beta^*$ the process also has an unexpected connection with a subclass of systems of linear equations, which we refer to as ``one-sink'' Laplacian systems. The importance of this subclass comes from the fact that the effective resistance between a pair of nodes in a network can be computed by solving a one-sink Laplacian system~\cite{Spielman-SICOMP:2011}~\cite{Madry-SODA:2015}. Over the last few years this connection to effective resistance has been repeatedly exploited to develop state-of-the-art algorithms for computing max flows in networks~\cite{Christiano-STOC:2011}\cite{Becchetti-AAMAS:2018}, random spanning trees of graphs~\cite{KelnerII-FOCS:2009}\cite{Madry-SODA:2015}, graph sparsification \cite{Spielman-SICOMP:2011}\cite{Lee-SIAMJC:2018}, and expander generation~\cite{Frieze-SIAMJC:2014}. The connection of our Data Collection Process to one-sink Laplacian systems opens up a new direction for the design of efficient distributed algorithms computing an array of important structures and quantities on graphs as we have shown in~\cite{Gillani-arXiv:2019}. However, efficient algorithms based on the Data Collection Process depend on fundamental mathematical properties of the process. Specifically, a stationary distribution must exist, and convergence towards it must be guaranteed in a reasonable time. This paper addresses those needs.

The rest of the paper is organized as follows. In Section \ref{sec:results}, we discuss our main results. In Section \ref{sec:ergodicity}, we prove existence of a non-trivial critical data rate below which the process is ergodic and above which it is non-ergodic. Then, in Section~\ref{sec:bound_rate} we characterize this rate in terms of underlying graph parameters. In Section~\ref{sec:geo_ergod}, we prove that the process is not only ergodic but geometrically ergodic and find the rate of convergence of the associated Markov chain to its stationary distribution. Finally, we conclude and give some directions for future work in Section \ref{sec:concl}.

\section{Main results}
\label{sec:results}
\subsection{Our model: The Data Collection Process}
\label{subsec:graph_data_coll_model}
We consider a stochastic process on a network modeled by an undirected graph $G=(V,E,w)$, where $V$ is the set of $n$ nodes, $E$ is the set of edges such that $|E|=m$, and a positive weight function $w:E\rightarrow \RR_+$ .  We say that $u \sim v$ if $(u,v) \in E$ and $\nbd{u} := \{v \in V | (u,v) \in E\}.$ The generalized degree of node $u$ is defined as $\degree{u} :=\sum_{v\in \nbd{u}}w_{uv}$. We denote the maximum and minimum generalized degree among all nodes in the network by $d_{\max}$ and $d_{\min}$ respectively. 

We consider time to be discrete and define the process in terms of the generation, movement and disappearance of ``packets'' from the system. In order to do this we are given a {\em relative rate vector} $\bm{J} \in \RR^n$ with the properties that (i) $\bm{J}(v) < 0$ for exactly one node and (ii) $\sum_{i=1}^n \bm{J}(i) = 0$. The node $v$ for which $\bm{J}(v) < 0$ is called the {\em sink} and we will use $\sink$ to denote it hereafter. We also define a set of {\em source nodes}: $V_s = \{v: \bm{J}(v) > 0\}$. We are also given a {\em rate parameter} $\beta \geq 0$ such that $\max_{i=1}^n \beta \bm{J}(i) \leq 1$. We assume that each node in $\nonsinks$ is equipped with a queue. The number of packets in the queue at $u$ at time $t$ is denoted by $Q_t(u)$. 


Packets appear in the system at the source nodes $v\in V_s$ which receive external packet arrivals as an independent Bernoulli process with rate $\beta\bm{J}(v)$. The packet received externally is placed in the queue at $v$. Packet movement at time $t$ takes place as follows: For each $u \in \nonsinks$, if $Q_t^\beta(u) >0$ a single data packet is picked at random from the queue and sent to $v$ with probability $w_{uv}/\degree{u}$. So, each node sends at most one packet from its queue in one time step and may receive multiple packets, up to one from each neighbour. A packet is removed from the system when a neighbor of $\sink$ decides to transmit that packet to $\sink$. 

In the following we will refer to the $|V|-1$-dimensional Markov chain $\left\{Q^{\bm{J},\beta}_t \right\}_{t\geq 0}$ as the {\em Data Collection Process on $G$ with relative rate vector $\bm{J}$ and rate parameter $\beta$}. Mostly we will omit $\bm{J}$ from the superscript since it will be understood. Occasionally we will maintain $\beta$ in the superscript but dispense with it when it is understood.

\subsection{Ergodicity is a critical phenomenon for the Data Collection process}
\label{subsec:process_steady_state}

The Data Collection process has two distinct regimes, one ergodic and one non-ergodic, as we vary $\beta$ and there is a sharp transition between them. We find that there is a non-trivial $\beta^* > 0$ such that the chain $\left\{Q^{\bm{J},\beta}_t \right\}_{t\geq 0}$ is ergodic for $\beta$ below this value and converges to a stationary distribution. Above $\beta^*$ the system displays drift and the queue sizes grow unbounded as $t \rightarrow \infty$. Specifically we show the following theorem:
\begin{theorem}
\label{thm:non_trivial_beta}
Consider a weighted undirected graph $G = (V,E,w)$ and a relative rate vector $\bm{J}$ with $\bm{J}(v) < 0$ for exactly one $v \in V$. If the random walk on $G$ with transition matrix $P_w$ where $\probm{u,v} = w_{uv}/\degree{u}$ is irreducible and aperiodic then there exists a $\beta^*>0$ such that the resulting multi-dimensional Markov chain $\left\{Q^{\bm{J},\beta}_t \right\}_{t\geq 0}$ is ergodic for all $\beta < \beta^*$ and non-ergodic for all $\beta\geq\beta^*$. 
\end{theorem}

Although it is difficult to prove ergodicity results for multi-dimensional Markov chains in general, we show in Section~\ref{sec:ergodicity} how the induction-based technique developed by Georgiadis and Szpankowski \cite{Georgiadis-QS:1992}, and later summarized by Szpankowski in his study of slotted ALOHA~\cite{Szpankowski-AAP:1994}, can be applied to prove this result.



\subsection{A lower bound on the critical rate}
\label{subsec:rate_result}
When $\beta < \beta^*$ the Data Collection process is ergodic and has a stationary distribution so we can define $\bm{\eta}^{\beta}(v) = \lim_{t \rightarrow \infty} \proby{Q_t^\beta(v) > 0}$ for all $v \in \nonsinks$. 
We will show in Section~\ref{sec:bound_rate} that at stationarity the vector $\bm{\eta}$ extended to $\sink$ by setting $\bm{\eta}(\sink) =0$ is a solution a linear system because 
\[ \bm{\eta}^T(I-P_w)= \beta \bm{J}^T,\]
where $P_w$ is the transition matrix of the random walk defined on $G$ by the weight function $w$. In Section~\ref{subsec:equiv_lap} we discuss the relationship of this system to the Laplacian of $G$ and the implications of this relationship. For now, we state one important consequence of this relationship: a lower bound on $\beta^*$.
\begin{theorem}
\label{thm:rate_bounds}
Suppose we have a Data Collection process  with relative rate vector $\bm{J}$ such that $\bm{J}(v) < 0$ only for $v = \sink$, defined on a graph $G=(V,E,w)$ that satisfies the conditions of Theorem~\ref{thm:non_trivial_beta} and has critical rate $\beta^*$. Then if $P_w$ is the transition matrix of the random walk defined by $w$ on $G$ and $\lambda_2^w$ is the second largest eigenvalue of $P_w$ then \begin{equation}
\beta^* \geq \frac{(1 - \lambda_2^w)}{\sum_{i\in V_s} \bm{J}(i)}~\frac{\sqrt{d_{\min}d_{\sink}}}{\left(d_{\max}+d_{\sink} \right)}.
\label{eq:lambda_lower_bounds_gen_srw}
\end{equation}
\end{theorem}

\subsection{Geometric Ergodicity}
\label{subsec:geo_ergod}
We show that when $\beta < \beta^*$ the Data Collection process converges to its stationary distribution at a geometric rate, i.e., the process is geometrically ergodic. Following Meyn and Tweedie \cite{Meyn-BOOK:1993}, we define geometric ergodicity formally:
\begin{definition}[Geometric ergodicity]
\label{def:geo_ergod}
Given an irreducible and aperiodic Markov chain $\Phi$ defined on state space $\mathcal{X}$ with transition probability $\mathcal{P}[\cdot,\cdot]$ and stationary distribution $\pi$, the chain is said to be {\em geometrically ergodic} if there exist constants $\rho < 1$, $R > 0$, and, for every state $\bm{x} \in \mathcal{X}$ there exists a $C_{\bm{x}} < \infty$, such that for all $t > 0$,
\begin{equation}
\| \probmt{\bm{x},\cdot} - \pi \| \leq R C_{\bm{x}} \rho^t.
\label{eq:geo_ergod_cond}    
\end{equation}
\end{definition}
We use the coupling method to prove that convergence happens at a geometric rate. The convergence rate is in terms of the {\em hitting time}, $\thit$, of the random walk $P_w$ defined on $G$ so we provide a definition of this quantity. If $\{X_t\}_{t \geq 0}$ is a random walk on $G$ and $\tau_v = \min \{t: X_t = v\}$, then $\thit = \max_{u,v \in V} \expt{\tau_v \mid X_0 = u},$ i.e., the maximum over all pairs $(u,v)$ of vertices of the expected time taken for a random walk begun at $u$ to first reach the vertex $v$. We show the following convergence theorem.
\begin{theorem}
\label{thm:geo_ergod_gen}
Consider $\left\{ Q_t^{\bm{J}, \beta}\right\}_{t\geq 0}$ defined on $G = (V,E,w)$ such that there is a critical $\beta^*$ as described in Theorem~\ref{thm:non_trivial_beta}. Let $\beta = \beta^*(1-\delta)$ for $\delta \in (0,1)$ and denote by $\mathcal{P}$ the transition matrix for the resulting multi-dimensional Markov Chain. Suppose we have $\bm{x},\bm{y} \in (\nn0)^{|V|-1}$ with $\sum_{i=1}^{|V|-1} \bm{x}(i) = N^{(\bm{x})}, \sum_{i=1}^{|V|-1} \bm{y}(i) = N^{(\bm{y})}$. Then 
\begin{equation}
||\probmt{\bm{x},\cdot}-\probmt{\bm{y},\cdot}||_{TV}\leq 2^{\left(\frac{8\thit+(8\max\{N^{(\bm{x})},N^{(\bm{y})}\}-1)\delta}{4\thit+\delta}\right)}\cdot \left(\frac{1}{2}\right)^{\frac{\delta}{4\thit+\delta}\cdot t }.
\label{eq:thm_rate_cong}
\end{equation}
\end{theorem}




Convergence to stationarity can be derived as a special case of Theorem~\ref{thm:geo_ergod_gen} by choosing $\bm{y} \in (\nn0)^{|V|-1}$ according to the $\pi$, the stationary distribution of chain $\left\{ Q_t^{\bm{J}, \beta}\right\}_{t\geq 0}$. This establishes the geometric ergodicity of the Data Collection process in the subcritical regime.

\begin{corollary}
\label{cor:geo_ergod_pi}
Consider the multi-dimensional Markov chain $\left\{ Q_t^{\bm{J}, \beta}\right\}_{t\geq 0}$ with $\beta = \beta^*(1-\delta)$ for $\delta \in (0,1)$ as defined in Theorem~\ref{thm:geo_ergod_gen} and denote its stationary distribution by $\pi$. For $\bm{x} \in (\nn0)^{|V|-1}$ such that $\sum_{i=1}^{|V|-1} \bm{x}(i) = N^{(\bm{x})}$, 
\begin{equation}
||\probmt{\bm{x},\cdot}-\pi||_{TV}\leq 4 \cdot 2^{\max \left\{\frac{8N^{(\bm{x})}\delta}{4\thit + \delta}, \frac{1}{2(1 - \delta)\beta^* + 1}\right\}}\cdot \left(\frac{1}{2}\right)^{\frac{\delta}{(4\thit+\delta)(2(1 - \delta)\beta^* + 1)} \cdot t}.
\label{eq:thm_geo_erg2}
\end{equation}
Moreover, for the special case that $\bm{x} = \bm{0}$, i.e., the system begins with empty queues, the Markov chain mixes to within $1/M$ of its stationary distribution in terms of total variation distance for any parameter $M > 0$ in time $t$ that is $\Theta\left(\frac{\thit \log M}{\delta}\right)$. 
\end{corollary}

\section{Ergodicity as a critical phenomenon}
\label{sec:ergodicity}
In this section, we prove existence of a non-trivial critical data rate $\beta^*$ for the multi-dimensional Markov chain $\left\{ Q_t^{\bm{J}, \beta}\right\}_{t\geq 0}$ associated with the Data Collection process such that the chain is ergodic for all values below $\beta^*$ and non-ergodic above it.

For a given a Data Collection process on a network modeled by an undirected graph $G=(V,E,w)$, there is an associated $\lvert V \rvert - 1$-dimensional vector $Q_t^\beta$ where each $Q_t^\beta(u)$ represents the queue size at a given node $u \in V\setminus \{\sink\}$ given a data rate $\beta$. Since the Data Collection process is a queueing system, the question of stability arises, i.e., we need to understand whether the system is able to successfully transfer data at a given value $\beta$ which is the controlling parameter for the rate at which packets appear in the system. For this, following Loynes \cite{Loynes-PCPS:1962} and Szpankowski~\cite{Szpankowski-AAP:1994}, we formally define a notion of a {\em stable data rate} as follows.
\begin{definition}[Stable rate]
Given a weighted undirected graph $G = (V,E,w)$ and a relative rate vector $\bm{J}$ with $\bm{J}(v) < 0$ for exactly one $v \in V$, the process  $Q^{\bm{J}, \beta}_t$ is said to be \emph{stable} and a value $\beta \geq 0$ of the rate parameter is said to be a {\em stable rate} if
\begin{equation}
\lim_{t \rightarrow \infty} \proby{\vert| Q^{\bm{J}, \beta}_t \vert|_{\infty} <x} = F(x), \mbox{ and }\lim_{x \rightarrow \infty} F(x) = 1
\label{eq:stab_cond}
\end{equation} 
where $F(x)$ is the limiting distribution function.
\end{definition}
However, if a weaker condition holds i.e., 
\begin{equation}
\lim_{x \rightarrow \infty}\lim_{t \rightarrow \infty}\inf \proby{\vert| Q^{\bm{J}, \beta}_t \vert|_{\infty} <x} = 1
\label{eq:sub_stable}
\end{equation}
the process is said to be \emph{substable} and otherwise \emph{unstable}. So, a stable process is necessarily substable and for a substable process to be stable its distribution function should tend to a limit. Thus, by stability we mean the distribution of $Q^{\bm{J}, \beta}_t$ as $t\rightarrow\infty$ exists. Moreover, if the limiting distribution is a \emph{stationary distribution}, then the process is \emph{ergodic}. So, for queueing systems ergodicity and stability can be used interchangeably.

In general, proving ergodicity of multi-dimensional Markov chain is difficult, however, for the Markov chain $\left\{ Q_t^{\bm{J}, \beta}\right\}_{t\geq 0}$ corresponding to our stochastic Data Collection process we can easily prove it.
This is because this process is part of a class of multi-queue systems for which Szpankowski and others showed a general method for proving the existence of a ``stability region'' of this kind~\cite{Szpankowski-AAP:1994}. Building on the work of Maly\v{s}ev~\cite{Malyshev-DAN:1972*} on two-dimensional Markov chains as extended by Maly\v{s}ev and Men\v{s}ikov~\cite{Malyshev:TMM:1981} to multi-dimensional chains, Georgiadis and Szpankowski developed an induction-based technique to characterize the stability region of the multi-queue system described by token passing rings~\cite{Georgiadis-QS:1992}. After applying this technique to several related systems, Szpankowski noted in his study of slotted ALOHA~\cite{Szpankowski-AAP:1994} that all the systems amenable to this technique had certain properties. We will first discuss that general characterization (properties) and then show that the Data Collection process falls within it. 

Given a multi-queue process $N_t$ with a set of $M$ queues. Let us consider a partition of $M$, $\mathscr{P}=(P,U)$ where $P$ refers to the set of {\em persistent} users which can transmit dummy packets even when their queues are empty and $U$ refers to the set of {\em non-persistent users} which behave as having normal queues. Now, for the given partition $\mathscr{P}$, let us define a modified multi-queue process  $\bar{N}_t^{\mathscr{P}}$ wherein the queues in $P$ are never allowed to become empty and queues in $U$ behave similar to those in original process $N_t$. To characterize the stability region of such processes like  $N_t$, Szpankowski's induction-based technique requires three conditions to hold:
\begin{enumerate}
\item {\em Monotonicity.} The queues in the modified process are always longer due to the persistent users, i.e., $N_t \preceq_{\mbox{\tiny SD}}\bar{N}_t^{\mathscr{P}}$.
\item {\em Stationarity of $U$.} Since, the users in set $U$ mimic the original process, the transmissions from $U$ that enter $P$ should form a stationary and ergodic sequence so that Loynes' scheme for one-dimensional queues~\cite{Loynes-PCPS:1962} 
can be applied to establish the stationarity of a persistent queue (in order to perform the induction step).
\item {\em Identical behaviors when non-empty.} $N_t$ and $\bar{N}_t^{\mathscr{P}}$ behave identically as long as their queues are non-empty. Only when $N_t(u)$ empties for some $u\in M$  and $\bar{N}_t^{\mathscr{P}}(u)$ is non-empty for $u\in P$, they behave differently.
\end{enumerate}

Szpankowski's general characterization is primarily based on an intrinsic coupling between the two processes $N_t$ and $\bar{N}_t^{\mathscr{P}}$ as indicated by the first and third property. In this coupling, starting from same initial state the transmission decisions are followed in the two processes i.e., if one process makes a transmission decision then the same decision is followed in the other, so that the trajectories of the two processes are coupled. Note that even if any queue in one of the processes is empty and the corresponding queue in the other is non-empty, any transmission decision of the latter will still be followed by the former although due to empty queue it will have no effect on its queue size or that of its neighbors. To show that the Data Collection process on graph $G=(V,E,w)$ also falls within the domain of this general characterization, we will also use different variations of this coupling for the corresponding Markov chain $Q_t^\beta$ over space $\{0,1\}^{V\times \mathbb{N}} \times \{0,1\}^{E\times \mathbb{N}} $.

To start with, using coupling based argument we will first prove an interesting property about the Data Collection process and its corresponding multi-dimensional Markov chain which satisfies Szpankowski's first condition about the monotonicity. In particular, we will show that for the Markov chain $Q_t^\beta$, the queue occupancy probability of a node $\proby{Q^\beta_t(u)>0}$ is an increasing function of $\beta$ for all $u \in V\setminus \{\sink\}$ and it is continuous for all $\beta<\beta^*$ where $\beta^*$ is the critical rate above which the queues are unstable and below which they are stable.

\begin{lemma}
\label{lem:queue_occup_incr_cont_fun}
Given an undirected graph $G=(V,E,w)$ running a Data Collection process. Let $Q^\beta_t$
represent the queues at time $t$ for all nodes $u \in V\setminus \{\sink\}$. Then, for all such nodes $\proby{Q^\beta_t(u)>0}$ is 
\begin{enumerate}
    \item an increasing function of $\beta$, and
    \item continuous for all $\beta<\beta^*$ where $\beta^*$ is the critical data rate such that  all data rates $\beta<\beta^*$ are stable and $\beta\geq\beta^*$ are unstable.
\end{enumerate} 
\end{lemma}
\begin{proof}
{\em (1).} To prove this property, we will first establish that the multi-dimensional Markov chain $Q_t^\beta$ is stochastically ordered i.e., stochastically larger initial states will produce stochastically larger chains at all times. For this, let us consider a coupling as used by Szpankowski of two trajectories of this chain $\{Q_t^\beta\}$ and $\{\bar{Q}_t^{\beta}\}$ such that $\bar{Q}_0^\beta \preceq_{\mbox{\tiny SD}} Q_0^\beta$. Now, assume the stochastic dominance relation between the two holds at time $t$ i.e., $\bar{Q}_t^\beta \preceq_{\mbox{\tiny SD}} Q_t^\beta$. Then, at time step $t+1$ for both $Q_t^\beta$ and $\bar{Q}_t^\beta$ from the one-step basic queue evolution equation at all nodes $u\in \nonsinks$ we have
\begin{align}
\expt{Q_{t+1}^\beta(u) \mid Q_t^\beta(u)} &= Q_t^\beta(u) - \ind{Q_t^\beta(u) > 0} \sum_{v:v \sim u} \probm{u,v} \nonumber\\ &+ \sum_{v:v \sim u} \probm{v,u} \ind{Q_t^\beta(v) > 0}
+A_t(u)
\label{eq:one_step_queue_eq_stab_proof}
\end{align} 
where $A_t(u)$ is the number of packets generated at $u$, which is 0 if $u \notin V_s$ and is 1 with probability $\beta \mathbf{J}(v)$ if $v \in V_s$, so, $\expt{A_t(u)}=\beta \mathbf{J}(u)$. Now consider any node $u$ at time $t+1$, from the induction hypothesis queues at node $u$ as well as its neighbors in $Q_t^\beta$ will dominate over the ones in $\bar{Q}^\beta_t$, so the first three terms on the right of Eq.~\eqref{eq:one_step_queue_eq_stab_proof} in $Q_t^\beta(u)$ will dominate the ones for $\bar{Q}^\beta_t(u)$ and since $\beta$ is same, the last term is same for both cases. So, we have $\bar{Q}_{t+1}^\beta(u) \preceq_{\mbox{\tiny SD}} Q_{t+1}^\beta(u)$. This is true for all nodes $u\in \nonsinks$, so we have at time $t+1$, $\bar{Q}_{t+1}^\beta \preceq_{\mbox{\tiny SD}} Q_{t+1}^\beta$. Hence, by induction the Markov chain $Q_t^\beta$ is stochastically ordered.

Now to prove monotonicity, for $\beta<\beta'$ let us consider a coupling similar to the one used before of two stochastically ordered Markov chains $Q_t^\beta$ and $Q_t^{\beta'}$ such that $Q_0^\beta \preceq_{\mbox{\tiny SD}} Q_0^{\beta'}$. Then, as we know for all $u\in \nonsinks$, $\beta \bm{J}(u)<\beta' \bm{J}(u)$, so by using induction and evolving queues using one-step queue evolution equation (Eq.~\eqref{eq:one_step_queue_eq_stab_proof}),  we can show that $Q_t^\beta \preceq_{\mbox{\tiny SD}} Q_t^{\beta'}$ for all $t$. Hence, by induction we have  $\proby{Q^\beta_t(u)>0}$ is an increasing function of $\beta$ for all $u\in \nonsinks$. 

{\em (2).} To prove the continuity of the given function for $\beta<\beta^*$, we will again consider a similar coupling, however between two stochastically ordered Markov chains $Q_t^\beta$ and $Q_t^{\beta-d\beta}$ with infinitesimal $d\beta$. For the data generation rule in the two chains, we have whenever new data packet is generated at any node in $Q^{\beta-d\beta}_t$ chain then, it is definitely generated at the corresponding node in $Q^{\beta}_t$ chain but not vice-versa. To understand the difference in the two chains, let $N^{\beta-d\beta}_t$ and $N^\beta_t$ denote the total number of packets in the respective chains till time $t$ and $\Lambda_t^{\beta}=N^\beta_t-N^{\beta-d\beta}_t$. Now, consider $g:[0,1]\rightarrow\RR$ to be a function dependent on $\beta$ such that $g(\beta)=\expt{Q_{t+1}^\beta(u)-Q_t^\beta(u)}$ which is bounded by definition. So, if we look at the derivative of this function, the term where  $\Lambda_t^{\beta}=0$ will be zero by definition of coupling, as the two chains behave differently only when there is an extra generated packet. Similarly, terms with $\Lambda_t^{\beta}\geq 2$ will have higher powers of $d\beta$ which will become zero as $d\beta\rightarrow 0$. Hence, the derivative $g'(\beta)$ only depends on $\Lambda_t^\beta= 1$ term i.e., $$g'(\beta)=\lim_{d\beta\rightarrow 0}\left(g(\beta)-g(\beta-d\beta)\mid\Lambda_t^\beta=1\right) \left(|V_s|t (1-d\beta)^{|V_s|t-1}\right)$$
where $V_s\subset V$ is the set of data sources. So, the total number of data packets generated in the two Markov chains upto time $t$ differ by one and hence, the queues at nodes in the two chains differ by at most one data packet at any time step. 
Now, for the given coupled chains let $t'$ be the time by which an extra packet is generated in chain $Q_t^\beta$. So, we have,
\begin{equation} 
\proby{Q_t^\beta(u)>0|Q^{\beta-d\beta}_t(u)=0}=\dfrac{\proby{Q_t^\beta(u)>0\cap Q^{\beta-d\beta}_t(u)=0}}{\proby{Q^{\beta-d\beta}_t(u)=0}} = \sum_{t'=1}^t d\beta P_{t'}
\label{eq:dbeta_exact}
\end{equation}
where $P_{t'}$ is the probability that the extra packet generated in chain $Q_t^\beta$
is present at node $u\in V\setminus\{\sink\}$. This means 
\begin{equation}
\proby{Q_t^\beta(u)>0}-\proby{Q_t^{\beta-d\beta}(u)>0}\leq\sum_{t'=1}^t d\beta P_{t'}.
\label{eq:extra_pkt_at_u}
\end{equation}

So, if $\proby{Q_t^\beta(u)>0|Q^{\beta-d\beta}_t(u)=0}$ is defined, as, $d\beta\rightarrow 0$ from the above equation we have, $\proby{Q_t^\beta(u)>0}-\proby{Q_t^{\beta-d\beta}(u)>0}\rightarrow 0$. Similarly, for the other side if  $\proby{Q^{\beta+d\beta}_t(v)>0|Q_t^{\beta}(u)=0}$ is defined, so as $d\beta\rightarrow 0$, similar to Eq.\eqref{eq:extra_pkt_at_u} we have, $\proby{Q_t^{\beta+d\beta}(u)>0}-\proby{Q_t^{{\beta}}(u)>0}\rightarrow 0$. Now, if both these conditions are true then the function is continuous as it has both left and right continuity respectively.

Now, consider all data rates $\beta<\beta^*$ where $\beta^*$ is the critical rate below which all rates are stable and above which all are unstable. So, for such rates both the probabilities $\proby{Q_t^\beta(u)>0|Q^{\beta-d\beta}_t(u)=0}$ and $\proby{Q^{\beta+d\beta}_t(v)>0|Q_t^{\beta}(u)=0}$ are defined, so as discussed above the function is continuous on both sides for all $\beta<\beta^*$. Now consider the case of data rates $\beta\geq\beta^*$. At $\beta^*$, we know $\proby{Q_t^{\beta^*}(u)>0}-\proby{Q_t^{\beta^*-d\beta}(u)>0}$ is defined (see Eq.~\eqref{eq:dbeta_exact}), as rate $\beta^*-d\beta$ is stable by definition, hence, the function is left continuous for this rate. However, for the other side since we know $\beta^*$ is not stable i.e., $\lim_{t\rightarrow\infty}\proby{Q^{\beta^*}_t(v)=0}=0$, hence, $\proby{Q^{\beta^*+d\beta}_t(v)>0|Q_t^{\beta^*}(u)=0}$ will not be defined and function is not right continuous. So, for $\beta\geq\beta^*$ function is left continuous but not right continuous. However, for all $u \in V\setminus \{\sink\}$, $\proby{Q^\beta_t(u)>0}$ is a continuous function (both limits exist) for all $\beta<\beta^*$.

\end{proof}

Having satisfied Szpankowski's first condition of monotonicity, we shall use two other general results to characterize the stability region of the multi-dimensional Markov chain associated with the Data Collection process. In particular, we will use Szpankowski's ``isolation lemma'' (Lemma~\ref{lem:isolation}) and Loynes' scheme~\cite{Loynes-PCPS:1962} as adapted to our situation (Lemma~\ref{lem:loynes_schm}).
\begin{lemma}[Szpankowski \cite{Szpankowski-ESP:1990*}]
\label{lem:isolation}
Given $N_t=(N_t^1,N_t^2,\cdots,N_t^M)$, an $M$-dimensional Markov chain.
\begin{enumerate}
    \item If it is defined on a countable state space, then the stability of $N_t^j$ for all $j\in M$ implies the stability of the multi-dimensional Markov chain $N_t$.
    \item If for some $j$, say $j^*$, $N_t^{j^*}$ is unstable, then $N_t$ is also unstable.
\end{enumerate}
\end{lemma}
\begin{lemma}[Loynes \cite{Loynes-PCPS:1962}]
\label{lem:loynes_schm}
Given a pair $(X_t^j,Y_t^j)$ of a strictly stationary and ergodic process, let $U_t^j=X_t^j-Y_t^j$. Then, the following holds:
\begin{enumerate}
    \item If $\expt{U_t^j}<0$, then $N_t^j$ is stable.
    \item If $\expt{U_t^j}>0$, then $N_t^j$ is unstable and $\lim_{t\rightarrow\infty} N_t^j=\infty$ (a.s.).
\end{enumerate}
\end{lemma}

Using these tools and Szpankowski's general method we will now prove the existence of a non-trivial stability region for the multi-dimensional Markov chain $\left\{ Q_t^{\bm{J}, \beta}\right\}_{t\geq 0}$ corresponding to a Data Collection process defined on an undirected graph $G=(V,E,w)$.

\begin{proof}[Proof of Theorem~\ref{thm:non_trivial_beta}] We first proceed by proving the sufficient part i.e., existence of a non-trivial $\beta^*>0$ such that the multi-dimensional Markov chain is ergodic for all $\beta<\beta^*$ and then the necessary part of the argument i.e., for all $\beta\geq\beta^*$ the chain is non-ergodic.

\paragraph{Sufficiency.} Given a partition $(P,U)$ of $\nonsinks$ queues we define a modification of $|V|-1$-dimensional chain $Q_t^\beta$ represented as $\bar{Q}_t^{\beta,U}$ where all nodes in $U$ have the same behavior as in $Q_t^\beta$ but the nodes in $\nonsinks \setminus U$ are not allowed to have empty queues.
Let us now first set $U = \emptyset$ (non-persistent users) and $P = \nonsinks$ (persistent users). For any $\beta \in (0,1)$, we know the one step basic queue evolution equation under the Data Collection process for any $u$ is as follows.
\begin{align*}
\expt{Q_{t+1}^\beta(u) \mid Q_t^\beta(u)} &= Q_t^\beta(u) - \ind{Q_t^\beta(u) > 0} \sum_{v:v \sim u} \probm{u,v} \\
&+ \sum_{v:v \sim u} \probm{v,u} \ind{Q_t^\beta(v) > 0} +\beta \bm{J}(u).
\end{align*}
So, at each node $u$ we have an arrival from $v$ with probability $\probm{v,u}$ in $\bar{Q}_t^{\beta, \emptyset}$ since the queue of $v$ is always non-empty and the departure is the usual $\sum_{v:v \sim u} \probm{u,v}$. 

Now, since we know $\probm{\sink, v}= 0$ for all $v \in \nonsinks$, so the sum of the outgoing probabilities from $\nonsinks$ is greater than the sum of the incoming probabilities, i.e., $\sum_{u \in \nonsinks}\sum_{v:v \sim u} \probm{u,v} > \sum_{u\in \nonsinks} \sum_{v:v \sim u,  v \in \nonsinks} \probm{v,u}.$ Therefore, there must be a vertex $u^* \in \nonsinks$ for which $\sum_{v:u^* \sim v} \probm{u^*,v} > \sum_{v:u^* \sim v, v \in \nonsinks} \probm{v,u^*}$. So, from Eq.~\eqref{eq:one_step_queue_eq_stab_proof} for this $u^*$ we note that the expected drift is
$$-\sum_{v:u^* \sim v} \probm{u^*,v} + \sum_{v:u^* \sim v, v \in \nonsinks} \probm{v,u^*} + \beta \bm{J}(u^*)$$
which is negative for an appropriately small but non-zero value of $\beta$, let's call it $\beta_{u^*}$. 

Now, to apply Loynes' scheme to vertex $u^*$ we need to ensure that the sequence $(I_t(u^*), O_t(u^*))$ is strictly stationary where $I_t(u^*)$ is the number of incoming packets to $u^*$ at time $t$ and $O_t(u^*)$ is the number of outgoing packets from $u^*$. Since all nodes $u \in P$, so $u^*$ as well as its neighbors always have a packet in the queue, so, both $O_t(u^*)$ and $I_t(u^*)$ are sequences of independent Bernoulli random variables and hence are stationary and ergodic. So, we can apply Loynes' scheme (Lemma~\ref{lem:loynes_schm}) to claim that the one-dimensional process $\bar{Q}^{\beta_{u^*}, \emptyset}_t(u^*)$ is stable, and, hence, $Q^{\beta_{u^*}}_t(u^*)$ is stable.

Now, we assume there is a non-empty set $U$ of non-persistent users and a $\beta_U > 0$ such that  $\bar{Q}_t^{\beta_U,U}(U)$ is stable and has a stationary distribution. To apply Loynes' scheme to a vertex, $u \in P  = \nonsinks \setminus U$ we need to ensure that the sequence $(I_t(u), O_t(u))$ is strictly stationary. Since $u \in P$ there is always a packet in the queue at $u$ and so $O_t(u)$ is a sequence of independent Bernoulli random variables which take value 1 with probability $\sum_{v:v\sim u} \probm{u,v}$ and 0 otherwise. We decompose $I_t(u)$ as the sum 0-1 random variables $A^{uv}_t$, where $A^{uv}_t = 1$ if $u$ receives a packet from $v$ at time $t$. Then 
\[I_t(u) = \sum_{v \in U} A^{uv}_t + \sum_{v \in P} A^{uv}_t.\]
Since all $v \in P$ have a packet in their queue at all $t \geq 0$, each $\sum_{v \in P} A^{uv}_t$ is the sum of Bernoulli random variables and hence taken from a strongly stationary sequence. If we start the $\bar{Q}_t^{\beta_U,U}$ from an initial state picked according to this stationary distribution which ensures that the process stays in the stationary state for all $t \geq 0$. In particular, this implies that for any $v \in P$, number of incoming packets from $v$ at time $t \geq 0$ is a sequence of random variables that is strongly stationary. Therefore $(I_t(u), O_t(u))$ is a strongly stationary sequence and we can apply Loynes' scheme. The expected drift at time $t \geq 0$ at any $u \in P$ for any $\beta \leq \beta_U$ is given by 
\begin{equation}
\label{eq:is-drift}
-\sum_{v:u\sim v} \probm{u,v} + \sum_{u \sim v, v\in P} \probm{v,u} + \sum_{u \sim v, v\in U} \probm{v,u}\probm{\bar{Q}^{\beta,U}_t(u) >0} +  \beta \bm{J}(u).
\end{equation}
Since the graph is connected and so there is at least one pair $(w_1,w_2)$ such that $w_1 \in U, w_2 \in P$ and $\probm{w_1, w_2} >0$, therefore we know that 
$\sum_{u \in P}\sum_{v \sim u} \probm{u,v} > \sum_{u\in P, v \in U} \sum_{v \sim u} \probm{v,u}.$
This means that there is a $u^* \in P$ such that  
$\sum_{u^* \sim v} \probm{u^*,v} > \sum_{u^* \sim v, v \in P} \probm{v,u^*}.$
For this $u^*$ the first two terms in Eq.~\eqref{eq:is-drift} add up to a value which is negative. Further from Lemma~\ref{lem:queue_occup_incr_cont_fun} we note that the third term is continuous and increasing in $\beta$ and tends to 0 as $\beta \downarrow 0$. Hence, it is possible to find a value $\beta_{U \cup \{u^*\}}$ which lies in $(0,\beta_U)$ such that the expected drift is negative. So, from Loynes' scheme (Lemma~\ref{lem:loynes_schm}) this implies that $\bar{Q}^{\beta, U}_t(U \cup \{u^*\})$ is stable for $\beta < \beta_{U \cup \{u^*\}}$. Moreover, from Lemma~\ref{lem:isolation} since the stability of all the one-dimensional Markov Chains associated with the vertices in $U \cup \{u^*\}$ implies the stability of the overall multi-dimensional chain. Consequently, the same holds for $Q_t^\beta(U \cup \{u^*\})$. Therefore by induction there is a $\beta^*$ such that for $\beta<\beta^*$, $Q_t^\beta$ is stable.

\paragraph{Necessity.} 
Corresponding to the sequence by which the stability region is expanded to include all the vertices of $\nonsinks$ there is a sequence $\beta_{u_1}, \beta_{u_2},\ldots, \beta_{u_{|\nonsinks|}}$ such that $\beta^{*} = \min \{\beta_{u_1}, \beta_{u_2},\ldots, \beta_{u_{|\nonsinks|}}\}.$ Let $w$ be the vertex for which $\beta_w = \beta^*$. Assume for the sake of simplicity of presentation that $\beta_w < \min \{\beta_u : u \in \nonsinks \setminus \{w\} \}.$ Hence we can choose any $\beta$ such that 
$\beta_w < \beta < \min \{\beta_u : u \in \nonsinks \setminus \{w\} \}.$
For this $\beta$ we know that $Q_t^{\beta, \nonsinks \setminus\{w\}}(\nonsinks \setminus\{w\})$ is stable. If we start this chain from its stationary distribution then the number of packets that are transmitted from $\nonsinks \setminus \{w\}$ to $w$ form a strongly stationary sequence. Since $w$ is persistent in this setting the packets leaving it are also strongly stationary. Hence Loynes' scheme (Lemma~\ref{lem:loynes_schm}) can be applied. By the choice of $\beta$ we know that the expected drift at $w$ is strictly positive and so $\bar{Q}_t^{\beta, \nonsinks \setminus\{w\}}(w)$ is unstable and hence by Lemma~\ref{lem:isolation}, $\bar{Q}_t^{\beta, \nonsinks \setminus\{w\}}$ is unstable.

In order to show that $Q_t^\beta$ is also unstable for this choice of $\beta$ we will show there is a coupling of $Q_t^\beta$ and $\bar{Q}_t^{\beta, \nonsinks \setminus\{w\}}$ with an appropriately chosen initial condition such that the two models behave {\em exactly} similarly. We know on the set of sample paths (of positive probability) on which the queue at $w$ remains strictly positive the two coupled models behave exactly similarly because the difference only arises if the queue at $w$ becomes 0 at time $t$, in which case $\bar{Q}_{t+1}^{\beta, \nonsinks \setminus\{w\}}(w)$ is automatically set to 1 since $w$ is persistent and $Q^\beta_{t+1}(w)$ remains 0. 
Now, we know that $\bar{Q}_t^{\beta, \nonsinks \setminus\{w\}}(w)$ is unstable, so when we start $\bar{Q}_t^{\beta, \nonsinks \setminus\{w\}}(\nonsinks \setminus\{w\})$ according to its stationary distribution and we set the queue at $w$ to 1, there is positive probability that this queue never reaches 0. So, for those cases $Q_t^\beta(w)$ behaves similarly as $\bar{Q}_t^{\beta, \nonsinks \setminus\{w\}}(w)$ i.e., it is unstable.
Therefore with these initial conditions $Q^\beta_t(w)$ is not substable since with positive probability  
$\lim_{t\rightarrow \infty} \proby{Q^\beta_t(w) > m}$, for all finite $m$. Hence, $Q^\beta_t(w)$ is unstable and, by Lemma~\ref{lem:isolation}, $Q^\beta_t$ is unstable for our choice of $\beta$ and, by the monotonicity of the process (see Lemma~\ref{lem:queue_occup_incr_cont_fun}), it is unstable for all choices of $\beta \geq \beta^*$.
\end{proof}

Having established the existence of a non-trivial critical data rate for the Markov chain $\left\{ Q_t^{\bm{J}, \beta}\right\}_{t\geq 0}$ of Data Collection process below which the chain is ergodic and above which it is non-ergodic, we will now characterize this critical rate. 

\section{Characterizing the critical rate}
\label{sec:bound_rate}
In section~\ref{sec:ergodicity}, we proved the ergodicity of Markov chain associated with the Data Collection process and showed that its stationary distribution exists. Now, in this section we will show that at steady-state Data Collection process is same as a special class of Linear equations which we call as the ``one-sink'' Laplacian system. Using this equivalence we will derive a lower bound on the critical rate. We will also discuss some common topologies in context of this result and show some tight examples. Lastly, we will also present an upper bound on the critical rate.
\subsection{Equivalence to one-sink Laplacian systems}
\label{subsec:equiv_lap}

The basic one step queue evolution equation under the Data Collection process for any node $u\in V$ is as follows.
\begin{align}
\expt{Q_{t+1}^\beta(u) \mid Q_t^\beta(u)} &= Q_t^\beta(u) - \ind{Q_t^\beta(u) > 0} \sum_{v:v \sim u} \probm{u,v} \nonumber\\
&+ \sum_{v:v \sim u} \probm{v,u} \ind{Q_t^\beta(v) > 0} + A_t(u), 
\label{eq:one_step_queue_eq_1}
\end{align}
where the second and third term on the right-hand side of the above equation represents the transmissions sent to and received from the neighbors respectively and $A_t(u)$ is the number of packets generated at $u$, which is 1 with probability $\beta \bm{J}(u)$ if $u \in V_s$, for the sink $\beta \bm{J}(\sink)=-\beta \sum_{v \in V_s}\mathbf{J}(v)$, and  for all other nodes $\bm{J}(u)=0$, where $u\notin \{V_s\cup\{\sink\}\}$.
Now, taking expectations on both sides of Eq.~\eqref{eq:one_step_queue_eq_1} and let $\bm{\eta}^\beta_t(u) = \proby{Q_t^\beta(u) > 0}$ be the queue occupancy probability of node $u$ and observing that $\expt{A_t(u)}=\beta \mathbf{J}(u)$, where $\bm{J}$ is the relative rate vector, we have
\begin{equation}
\expt{Q_{t+1}^\beta (u)} = \expt{Q_t^\beta (u)}  - \bm{\eta}_t^\beta(u) \sum_{v:v\sim u}\probm{u,v} + \sum_{v:v\sim u}\probm{v,u}\bm{\eta}_t^\beta(v)+\beta \mathbf{J}(u).
\label{eq:exp_one_step_queue_eq}
\end{equation}

From Theorem~\ref{thm:non_trivial_beta}, we know that for an appropriately chosen value of $\beta$ the Data Collection process has a steady state. Moreover, at steady state $\expt{Q_t^\beta(u)}$ is a constant, so if we let $\bm{\eta}^{\beta}(u) = \lim_{t \rightarrow \infty} \proby{Q_t^\beta(u) > 0}$ be the queue occupancy probability of node $u$ at the stationarity, then we have the steady-state equation for the given node as
\begin{equation}
- \bm{\eta}^\beta(u) \sum_{v:v\sim u}\probm{u,v}  + \sum_{v:v\sim u}\probm{v,u}\bm{\eta}^\beta(v)+\beta \mathbf{J}(u)= 0.
\label{eq:steady_state_queue_eq_gen}
\end{equation}
We can also represent the steady-state equations of all $|V|=n$ nodes in matrix form as follows. For this, let us first order the nodes such that the $n$th node represents the sink. Let $\bm{\eta}$ be an $n$ element column vector representing the steady-state queue occupancy probability $\bm{\eta}^\beta(u)$ of nodes $u\in V$. We drop the subscript $\beta$ where the rate is understood from the context. So, we have $\bm{\eta} = [\bm{\eta}(1)~\bm{\eta}(2)~\cdots~\bm{\eta}(n-1)~0]$. This is defined assuming that sink collects all data it receives and has no notion of maintaining queue. Let $\bm{J}$ be another $n$ element column vector such that $\bm{J}(i)>0$ if $i\in V_s$, $\bm{J}(\sink)=-\sum_{i\in V_s}\bm{J}(i)$ and 0 elsewhere, and $I$ be the usual $n\times n$ identity matrix. So, given the transition matrix $P_w$ for the random walk defined by $w$ on graph $G$, the steady-state queue equations at the nodes can be written in matrix form as
\begin{equation}
\bm{\eta}^T(I-P_w)= \beta \bm{J}^T.
\label{eq:vector_queue_eq_1}
\end{equation}
As we know transition matrix $P_w=D^{-1}A$ where $D$ is the diagonal matrix of generalized degrees and $A$ is the adjacency matrix, so matrix $(I-P_w)$ is also a Laplacian as we can rewrite it as $(I-P_w)=D^{-1}(D-A)=D^{-1}L$. So, the above equation (Eq.~\eqref{eq:vector_queue_eq_1}) can be rewritten as 
\begin{equation}
\bm{x}^TL=\beta \bm{J}^T
\label{eq:steady-state_pot}
\end{equation}
 where $\bm{x}^T=\bm{\eta}^TD^{-1}$ is a row vector such that $\bm{x}(u)=\bm{\eta}(u)/\degree{u}$ for all $u$ where $\bm{\eta}(u)$ is the steady-state queue occupancy probability and $\degree{u}=\sum_{v:(u,v)\in E}w_{uv}$ is the generalized degree of node $u$. Eq.~\eqref{eq:steady-state_pot} is similar to Laplacian systems of the form $L\mathbf{x} = \mathbf{b}$ with a constraint that only one element in $\mathbf{b}$ is negative. We call such systems ``one-sink'' Laplacian systems. In our subsequent work \cite{Gillani-arXiv:2019} we discuss this connection in detail.
 
\subsection{A lower bound}
\label{subsubsec:lower_bound_beta}
Now having established the steady-state equation for the Data Collection process, we will use it for characterising the critical data rate. In particular, we will prove a lower bound on such rate.
\begin{proof}[Proof of Theorem \ref{thm:rate_bounds}]
For a given graph $G=(V,E,w)$, with source set $V_s\subseteq V\setminus\{\sink\}$ and transition matrix $P_w$ for random walk defined by $w$ on graph $G$, 
recall that the steady-state queue equations at nodes 
can be written in vector form as 
\begin{equation}
\bm{\eta}^T(I-P_w)=\beta \bm{J}^T.
\label{eq:matrix_queue_eq}
\end{equation}

Now, in order to bound the maximum stable data rate $\beta$ at which the source nodes generate data in terms of the underlying graph parameters, we will consider eigendecomposition of the left hand side of Eq.~\eqref{eq:matrix_queue_eq}. For this, we will deviate from the usual inner product on the vector space $\mathbb{R}^V$ i.e., $\langle f,g\rangle = \sum_{x\in V}f(x)g(x)$ and define another inner product on $\mathbb{R}^V$ which is given by $\langle f,g\rangle_\mu := \sum_{x\in V}f(x)g(x)\mu(x)$ where $\mu$ is the stationary distribution of random walk defined by $w$ on graph satisfying $\mu=\mu P_w$. From Lemma 12.2 \cite{Levin-BOOK:2009},
it is known that the inner product space $(\mathbb{R}^V, \langle\cdot,\cdot\rangle)_\mu$ has an orthonormal basis of real-valued eigenfunctions $\{f_j\}_{j=1}^{|V|}$ corresponding to real eigenvalues $\{\lambda_j\}$.
Using this lemma and writing the vector $\bm{\eta}^T$ in terms of the eigenvectors, we have
$\bm{\eta}^T = \sum_{i=1}^{|V|} \langle\bm{\eta}^T,f_i\rangle_\mu f_i.$
This gives us that 
$\bm{\eta}^T(I - P_w) = \sum_{i=1}^{|V|} (1 - \lambda_i^w)\langle\bm{\eta}^T,f_i\rangle_\mu f_i$, where $\lambda_i^w$ is the $i^{th}$ eigenvalue of transition matrix $P_w$. Moreover, from Lemma 12.1 of \cite{Levin-BOOK:2009}, 
we also know that the absolute value of any eignevalue of a transition matrix can be at most $1$, so, $\lambda_1^w = 1 > \lambda_2^w \geq \cdots \geq \lambda_n^w$. So, we have
\begin{align}
\bm{\eta}^T(I - P_w) &= \sum_{i=2}^{|V|}(1 - \lambda_i^w)\langle\bm{\eta}^T,f_i\rangle_\mu f_i \\
 &\geq (1 - \lambda_2^w)\left(\sum_{i=2}^{|V|}\langle\bm{\eta}^T,f_i\rangle_\mu f_i\right).
\label{eq:mu_p_bound}
\end{align}
Note, that $f_1, \ldots, f_{|V|}$ form an orthonormal basis so,
$\sum_{i=1}^{|V|}\langle\bm{\eta}^T,f_i\rangle_\mu^2 = \lVert\bm{\eta}^T\rVert_\mu^2$. Hence we have  
\begin{equation}
\sum\limits_{i=2}^n{\langle\bm{\eta}^T,f_i\rangle_\mu^2} = \lVert\bm{\eta}^T\rVert_\mu^2 - \langle\bm{\eta}^T,f_1\rangle_\mu^2.
\label{eq:inner_product_i_2}
\end{equation}
The eigenfunction $f_1$ corresponding to the eigenvalue 1 can be taken to be a constant vector \textbf{1}, so $\langle\bm{\eta}^T,f_1\rangle_\mu = \sum\limits_{i=1}^n{\bm{\eta}(i)\mu(i)}$, where $\mu(i)=\sum_{v\in V} \mu(v)P_w[v,i]$ . Also, $\lVert\bm{\eta}^T\rVert_\mu^2 = \sum\limits_{i=1}^n{\bm{\eta}^2(i)\mu(i)}$. So, using these results in Eq. \eqref{eq:inner_product_i_2} we have
\begin{equation}
\sum\limits_{i=2}^n{\langle\bm{\eta}^T,f_i\rangle_\mu^2} = \sum\limits_{i=1}^n{\bm{\eta}^2(i)\mu(i)} - \Bigg(\sum\limits_{i=1}^n{\bm{\eta}(i)\mu(i)}\Bigg)^2 = \Var_\mu(\bm{\eta}(i)) = \sum\limits_{i=1}^n(\bm{\eta}(i) - \bar{\bm{\eta}}_\mu)^2\mu(i)
\label{eq:inner_product_i_2_expand}
\end{equation}
where, $\bar{\bm{\eta}}_\mu=\sum\limits_{i=1}^n{\bm{\eta}(i)\mu(i)}$ is the expected queue occupancy probability of nodes under stationary distribution $\mu$. Now, taking square of norm of Eq.~\eqref{eq:mu_p_bound} and using Eq.~\eqref{eq:inner_product_i_2_expand}, we have 
\begin{equation}
\lVert \bm{\eta}^T(I - P_w) \rVert_\mu^2 \geq (1 - \lambda_2^w)^2 \Var_\mu(\bm{\eta}(i)).
\label{eq:i-p_norm_2}
\end{equation}

Using Eq.~\eqref{eq:i-p_norm_2} in the square of norm of Eq. \eqref{eq:matrix_queue_eq}, we have
\begin{equation}
\beta \geq \frac{(1 - \lambda_2^w)}{\lVert \bm{J}^T \rVert_\mu}~\sqrt[]{\Var_\mu(\bm{\eta}(i))}.
\label{eq:gen_beta_J_var_term}
\end{equation}

Moreover, as $\sum_{i\in V_s}\bm{J}^2(i)\leq (\sum_{i \in V_s}\bm{J}(i))^2$, so we have
\begin{align}
\lVert \bm{J}^T \rVert_\mu&=\sqrt{\sum_{i \in V_s}\bm{J}^2(i) \mu(i) + \left(\sum_{i \in V_s} \bm{J}(i)\right)^2\mu(\sink) }
\leq \sum_{i \in V_s} \bm{J}(i)\sqrt{ \mu_m + \mu(\sink)}
\label{eq:denom_srw_j}
\end{align}
where $\mu_m=\max_{i \in V_s} \mu(i)$

Now to get a bound on $\Var_\mu(\bm{\eta}(i)) = \sum\limits_{i=1}^n(\bm{\eta}(i) - \bar{\bm{\eta}}_\mu)^2\mu(i)$, we consider two nodes whose queue occupancy probability we know precisely (1) the sink, $\sink$, which has $\bm{\eta}(\sink) = 0$ (as it has no notion of maintaining queue and it sinks data packets as soon as it receives them), and (2) a node $u_{\max}$ with maximum queue occupancy probability for a given $\beta$, let it be $\bm{\eta}_{\max}^\beta=\max_{u\in\nonsinks}\bm{\eta}^\beta(u)$. Now, let $\beta=(1-\delta)\beta^*$ where $\beta^*$ is the critical data rate and $\delta \in (0,1)$. From Eq.~\eqref{eq:vector_queue_eq_1} we know $\bm{\eta}$ is linear in $\beta$ and $\bm{\eta}_{\max}^{\beta*}=1$, so $\bm{\eta}_{\max}^\beta=\frac{\beta}{\beta^*}$ and hence, we have $\bm{\eta}_{\max}^\beta=1-\delta$.

We note that the contribution of $\sink$ and $u_{\max}$ with $\bar{\bm{\eta}}_\mu=\sum_{i=1}^n \bm{\eta}(i)\mu(i)$ as the expected queue occupancy probability of nodes under the stationary distribution $\mu$ is as follows.
\begin{equation}
\Var_\mu(\bm{\eta}(i)) \geq \left(1-\delta- \bar{\bm{\eta}}_\mu)^2\mu(u_{\max}) + (\bar{\bm{\eta}}_\mu - 0)^2\right)\mu(\sink) \geq \frac{(1-\delta)^2\mu(u_{\max})\mu(\sink)}{\mu(u_{\max})+\mu(\sink)}.
\label{eq:var_bound_pi}
\end{equation}
where the last inequality holds as $\left(1- \bar{\bm{\eta}}_\mu)^2\mu(u_{\max}) + (\bar{\bm{\eta}}_\mu - 0)^2\right)\mu(\sink)$ achieves optimum at $\bar{\bm{\eta}}_\mu=\frac{(1-\delta)\mu(u_{\max})}{\mu(u_{\max})+\mu(\sink)}$. So, first using Eq.~\eqref{eq:denom_srw_j} and Eq.~\eqref{eq:var_bound_pi} in Eq.~\eqref{eq:gen_beta_J_var_term} and then we know as $\beta \rightarrow \beta^*$, $\delta\rightarrow 0$, so we have
\begin{equation}
\beta^* \geq \frac{(1 - \lambda_2^w)}{\sum_{i\in V_s} \bm{J}(i)}~\sqrt{\frac{\mu_{u(\max)}\mu(\sink)}{(\mu(u_{\max})+\mu(\sink))(\mu_m+\mu(\sink))}}.
\label{eq:beta_final_pi}
\end{equation}

Now, we know $\mu(i)=\frac{\degree{i}}{\sum_{u\in V}\degree{u}}$, and $\frac{d_{\min}}{\sum_{u\in V}\degree{u}} \leq \mu(i) \leq \frac{d_{\max}}{\sum_{u\in V}\degree{u}}$ where, $d_{\min}$ and $d_{\max}$ are the generalized minimum and maximum degrees of graph respectively. 
So using the appropriate bounds on $\mu(i)$ in Eq.~\eqref{eq:beta_final_pi} we have
\begin{equation}
\beta^* \geq \frac{(1 - \lambda_2^w)}{\sum_{i\in V_s} \bm{J}(i)}~\frac{\sqrt{d_{\min}d_{\sink}}}{\left(d_{\max}+d_{\sink} \right)}
\label{eq:beta_final_degree_sink}
\end{equation}
where $\lambda_2^w$ is the second smallest eigenvalue of the transition matrix of random walk defined by weight function $w$ and $d_{\sink}$ is the generalized degree of the sink node.
\end{proof}

\begin{table}[!]
\caption[Rate bounds for various graphs]{Rate bounds for various graphs with $w:E\rightarrow \bm{1}$, $|V_s|=1$ such that $\sum_{i\in V_s} \bm{J}(i)=1$ }
\label{table:rate_bounds}
\begin{center}
\scalebox{0.9}{
 \begin{tabular}{lcc}
\hline
\\
\textbf{Graph} & $\beta \geq \frac{(1 - \lambda_2^w)}{\sum_{i\in V_s} \bm{J}(i)}~\frac{\sqrt{d_{\min}d_{\sink}}}{\left(d_{\max}+d_{\sink} \right)}$ &\textbf{Exact rate} \\
\hline
\\
Cycle  &$\dfrac{1}{2n^2}$ &$\dfrac{2}{n}$ \\
Star Graph with sink at centre & &\\
and $\epsilon$ as self loop probability at each node  &$\dfrac{1}{2\sqrt{n-1}}$ &$1-\epsilon$ \\
Star Graph with sink and source & &\\
at outer node  &$\dfrac{1}{n}$ &$\dfrac{1}{n-1}$ \\
\\
Complete graph  &$\dfrac{n}{2(n-1)}$ &$\dfrac{n}{2(n-1)}$ \\
\\
Random Geometric Graph  &$\dfrac{\log n}{2n}$ &- \\
Wheel Graph $W_{n+1}$ with sink at centre & &\\
and source at one of the cycle vertices  &$\dfrac{\log n\sqrt{3n}}{2n^2}$ &$\dfrac{1}{3}$ \\
Wheel Graph $W_{n+1}$ with source at centre & &\\
and sink at one of the cycle vertices  &$\dfrac{3\log n}{n(n+1)}$ &$\dfrac{5}{3n}$ \\
Complete Binary tree with both & &\\
source and sink at leaves  &$\dfrac{1}{4n}$ &$\dfrac{1}{6\log n-3}$ \\
$k$-times star of star graph & &\\
 with both source and sink at leaves  &$\dfrac{1}{n^2+n^{\frac{2k-1}{k}}}$ &$\dfrac{1}{1+(2k-1)n^{1/k}}$ \\

$k$-times star of star graph & &\\
 with source at center and sink at leaf  &$\dfrac{1}{2n^{\frac{4k-1}{2k}}}$ &$\dfrac{1}{1+(k-1)n^{1/k}}$ \\
\hline
\end{tabular}}
\end{center}
\end{table}
In Table~\ref{table:rate_bounds}, we present lower bound on the critical data rate for the stochastic Data Collection process. We also present the exact values of data rate which are easy to calculate using elementary algebra for these topologies. In all these cases, we assume that all edges have unit weight $w:E\rightarrow \bm{1}$ i.e., random walk defined by $P_w$ is simple random walk, there is only one source node i.e., $|V_s|= 1$ such that $\sum_{i\in V_s} \bm{J}(i)=1$.  

If we consider the complete graph topology it is easy to see that the exact rate is $n/2(n-1)$. As, the spectral gap of the simple random walk on the complete graph of $n$ nodes is $n/n-1$, we note that for this case our lower bound is tight i.e., both the exact value and the lower bound have order $\Theta(1)$. Similarly, for the star graph with sink at outer edge, our lower bound is tight and is of order $\Theta(1/n)$. Hence it is clear that our lower bound cannot admit any asymptotic improvement in general. On the other hand, consider cycle topology which shows that for specific cases a better lower bound may be possible. We note that our spectral gap-based lower bound is a $\Theta(1/n)$ lower than the exact value for this case. Similarly, for other topologies like wheel graph, complete binary tree and $k$-times star of star graph ($n^{1/k}$-regular tree defined on $k$ levels) a better lower bound is possible. 


\subsection{An upper bound}
\label{subsec:upper_bound}

We also prove an upper bound on the critical data rate for a special case where $V_s=\nonsinks$. In order to present this bound, we need to define some terms. For any vertex $u \in V$, we define its measure as, $\rho(u):=\sum\limits_{v\in V}\probm{u,v}$. Similarly, for any $U \subset V$ we define the measure $\rho(U)=\sum\limits_{u\in U}\rho(u)$. We also define the edge boundary as $\partial U:=\{(u,v): u \in U ,v \notin U\}$, so, $\rho(\partial U)= \sum\limits_{u \in U ,v \notin U}\probm{u,v}$. We have the following upper bound result.

\begin{proposition}
\label{pro:rate_upper_bound}
Given a graph $G=(V,E,w)$ with $|V|=n$ nodes out of which there is one sink $\sink$ and set $V_s = \nonsinks$ of source nodes running a Data Collection process having critical data rate $\beta^*$ as defined by Theorem~\ref{thm:non_trivial_beta}. To achieve stable queues $\beta^*$ must satisfy
\begin{equation}
\beta^* \leq \min\Big\{\hat{h}(G), \sum\limits_{u:u\sim \sink}\frac{\probm{u,\sink}}{n-1}\Big\}
\label{eq:ubound}
\end{equation}
 where $P_w$ is the transition matrix of random walk defined by $w$, $\hat{h}(G)=\min\limits_{U \subset V,\sink\notin U}\frac{\rho(\partial U)}{\rho(U)}$ is a constant and $\hat{h}(G)$ is at most $h(G)$, the edge expansion of graph $G$.
\end{proposition}
\begin{proof}[Proof of Proposition~\ref{pro:rate_upper_bound}]
Given any vertex $u \in V$, recall its measure is defined as, $\rho(u):=\sum\limits_{v\in V}\probm{u,v}$, and for any $U \subset V$ we have $\rho(U)=\sum\limits_{u\in U}\rho(u)$. Similarly, for edge boundary as $\partial U:=\{(u,v): u \in U ,v \notin U\}$, we have $\rho(\partial U)= \sum\limits_{u \in U ,v \notin U}\probm{u,v}$. Now, let us define constants $h(U):= \frac{\rho(\partial U)}{\rho(U)}$ and $\hat{h}(G):=\min\limits_{U \subset V,\sink\notin U}h(U) \leq h(G)$ where $h(G)$ is the edge expansion of graph $G$. 

We know, for any given set $U \subset V$, where $\sink\notin U$ the maximum data flow that can move out of this set is the flow across the boundary $\partial U$, so 

\begin{align}
\beta \rho(U) &\leq \rho(\partial U)\label{eq:flow_boundary}\\
\beta &\leq \min\limits_U h(U) = \hat{h}(G) \leq h(G)\label{eq:lambda_h(U)}
\end{align}

Now, for set $U =V_s=V\setminus \{\sink\}$, we have $\hat{h}(G))\leq \sum\limits_{u:u\sim \sink}\frac{\probm{u,\sink}}{n - 1}$. So, from eq.~\eqref{eq:flow_boundary} $\beta \leq \sum\limits_{u:u\sim \sink}\frac{\probm{u,\sink}}{n - 1}$. Hence, the upper bound on the critical data rate is given by,
\begin{equation}
\beta \leq \min\Big\{\hat{h}(G), \sum\limits_{u:u\sim \sink}\frac{\probm{u,\sink}}{n-1}\Big\}
\label{eq:lambda_upper}
\end{equation}
\end{proof}
Note that our derived upper and lower bound on the critical data rate relates directly to the two sides of Cheeger's inequality~\cite{Cheeger-PCHB:1969*}.

\section{Geometric rate of convergence}
\label{sec:geo_ergod}
Next, we characterize the rate of convergence of Markov chain $\left\{ Q_t^{\bm{J}, \beta}\right\}_{t\geq 0}$ for the stable regime i.e., $\beta< \beta^*$. In particular, we first prove a general result about the total variation distance between the probability distributions of two Markov chains and their rate of convergence. Then, as a special case of this result we show that the convergence of Markov chain $\left\{ Q_t^{\bm{J}, \beta}\right\}_{t\geq 0}$ is geometric i.e, starting from any initial state, the distance from the stationarity reduces exponentially. Note that we drop the superscript $\bm{J},\beta$ from the Markov chain representation as a stable data rate value for proving the convergence rate is assumed. 

\begin{proof}
[Proof of Theorem~\ref{thm:geo_ergod_gen}]
We first note that our Markov chain $Q_t$ is {\em stochastically ordered} (c.f.~\cite{Lund-MOR:1996}). In general this means, if we are given two random processes $X$ and $Y$ supported on $\nn0^{|\nonsinks|}$ we say $X$ is stochastically dominated by $Y$ if $\expt{f(X)}\leq \expt{f(Y)}$ for every increasing function $f$. For our Data Collection chain we state the stochastic orderedness property as follows.
\begin{claim}
\label{clm:stochastic-ordering}
Given two instances of the Data Collection process $Q_t$ and $Q'_t$ such that $Q_0 \preceq Q'_0$, $Q_t$ is stochastically dominated by $Q'_t, t \geq 0$. In particular this means that $\proby{Q_t(v) > 0} \leq \proby{Q'_t(v) > 0}$ for all $v \in \nonsinks$.
\end{claim}
The proof of this claim follows by constructing a coupling between the two chains such that each of them perform exactly the same transmission actions. In case one of the chains is empty then the transmission action is a dummy action. It is easy to see that stochastic ordering follows naturally for the Data Collection chain. 

To use this claim, for our irreducible and aperiodic Markov chain $Q_t$ described by the Data Collection process defined on $(\nn0)^{|V|-1}$ having transition matrix $\pmd$ and a stationary distribution $\pi$, let us define two other irreducible and aperiodic Markov chains $Q^1_t$ and $Q^2_t$, each with state space $(\nn0)^{|V|-1}$. Initially, suppose the data is generated in the two chains in a coupled way such that one of them dominates the other i.e., either $Q^1_0(v) \leq Q^2_0(v)$ for all $v \in \nonsinks$ or vice-versa. 

Now, consider the coupling $(Q^1_t,Q^2_t)$ on $(\nn0)^{|V|-1} \times (\nn0)^{|V|-1}$ defined over random sequences $\{0,1\} \times \{\prod_{v \in \nonsinks} \Gamma(v)\} $ where $\Gamma(v)$ is the set of one-step destinations from node $v$, such that both the chains $Q^1_t$ and  $Q^2_t$ are populated in a coupled way. Such Markov chains are said to be stochastically ordered chains  in the queueing theory and have a property that the Markov chain which dominates the other chain will always maintain dominance over it. 

Now, under this coupling we allow the two chains to run in a way that any data generation or data transmission decision made by any queue in one chain is followed by the corresponding queue in the other chain as well. However, to distinguish the newly generated packets in two chains from the existing ones, we assign colors to the data packets: the existing packets in $Q^1_t$ chain are colored red and in $Q^2_t$ chain are colored blue, and the newly generated packets in both the chains are colored green. Moreover, in both the chains green (newly generated) packets get a preference in the transmission. Now, let $Q_t^{1,green}(u)$ represent the number of green packets in the queue of a given node $u$ in $Q^1_t$ and $\bm{\eta}(u)$ be the steady-state queue occupancy probability of Markov chain $Q_t$. Since, the number of green packets in both the chains starts from zero and the chains are stochastically ordered, green packet queue occupancy is always bounded by that of the chain with stationary distribution i.e., $\proby{Q_t^{1,green}(u)\geq 1}\leq \bm{\eta}(u)$. Same holds true for the other chain $Q^2_t$ as well.  

Now, to ensure both chains get coupled all the red and blue (old) packets in $Q^1_t$ and $Q^2_t$ respectively need to be sunk. We consider $Q^1_t$ chain and the same will hold for $Q^2_t$ as well. We know by our preference in transmission, the probability that red packets move out of queue in one time step in $Q^1_t$ is equal to the probability that there are no green packets in the given queue i.e., $1-\proby{Q_t^{1,green}(u)\geq 1}$. Also, we have  $1-\proby{Q_t^{1,green}(u)\geq 1}\geq 1-\bm{\eta}(u)\geq \min_u 1-\bm{\eta}(u)\geq 1-\bm{\eta}_{max}$, where $\bm{\eta}_{max}=\max_{u\in\nonsinks}\bm{\eta}(u)$. Now, let $N^{(red)}$ and $N^{(blue)}$ be the total number of red and blue data packets in chains $Q^1_t$ and $Q^2_t$ respectively at the beginning which are assumed to be finite. Also, let $T_{N^{(red)}}$ and $T_{N^{(blue)}}$ be the time taken by the the respective number of packets to get sunk. We have the following lemma that bounds this time.
\begin{lemma}
\label{lem:old_pkt_sinking_time}
Given a Data Collection process on graph $G$ with $N^{(*)}<\infty$ as the total number of data packets present in the queues of all nodes initially, then the time taken by all packets to reach the sink, let it be $T_{N^{(*)}}$ is bounded as
\begin{equation}
\proby{T_{N^{(*)}}\geq \frac{4\thit}{1-\bm{\eta}_{\max}} (\log 1/\epsilon+2)+8N^{(*)}+\log 1/\epsilon-1}\leq \frac{\epsilon}{2}
\label{eq:tni_bound}
\end{equation}
where $\thit$ is the worst-case hitting time of random walk on $G$ and $\bm{\eta}_{\max}$ is the maximum queue occupancy probability at stationarity.
\end{lemma}
\begin{proof}[Proof of Lemma \ref{lem:old_pkt_sinking_time}]
To prove this lemma we follow the delay analysis by Leighton et al. \cite{Leighton-JA:1994}. So, for our given Data Collection process on graph $G$ with $N^{(*)}<\infty$ as the total number of data packets present in the queues of all nodes initially, each data packet has its own trajectory or trace of random walk which indicates its path to reach the sink. Moreover, to each of these 
$N^{(*)}$ packets we assign distinct ranks out of range $K$ which will be determined later and the packet with the lowest rank always gets preference in the transmission. Among all possible sets of ranks assigned to the packets we choose a particular trace of random walk and find a \emph{delay sequence} for it.

A delay sequence of length $s$ as defined by Leighton et al. involves backtracking the path of $s+1$ data packets where $s$ is determined by the analysis. In particular, given a data packet $p_1$ which arrived at the sink we follow it backwards till the edge it got delayed last time, suppose that edge is $e_1$. Let $\ell_1$ be the length of the path from the sink to edge $e_1$ and suppose $p_1$ got delayed by packet $p_2$. Then, we follow $p_2$ backwards till the edge where it got delayed by some packet. This is repeated till we get packet $p_{s+1}$ delayed packet $p_s$ over edge $e_s$. So, the path from $e_s$ to the sink forms a delay sequence. Moreover, the intermediate $s$ paths of length $\ell_1, \ell_2, \cdots, \ell_s\geq 0$ have the property that $\sum_{i=1i}^s \ell_i \leq D $ where $D$ is the maximum number of edges that can be traversed by a trace of random walk.

Now, from our earlier argument we know that probability that any of the $N^{(*)}$ (old) packet moves out of queue in one time step is at least  $1-\bm{\eta}_{\max}$, where $\bm{\eta}_{\max}$ represents the maximum queue occupancy probability at stationarity. This means any one step in this stochastic process takes on an average $\frac{1}{1-\bm{\eta}_{\max}}$ time. So, the expected time taken by any random walk to hit the sink is $\frac{\thit}{1-\bm{\eta}_{\max}}$ where $\thit$ is the worst-case hitting time of random walk. So, by Markov's inequality $\proby{D\geq \frac{2\thit}{1-\bm{\eta}_{\max}}} \leq\frac{1}{2}$.
Now, consider the probability of a random walk not hitting sink $\sink$ in $2(\log 1/\epsilon+2)$ times $\frac{\thit}{1-\bm{\eta}_{\max}}$ i.e., we consider $\frac{\thit}{1-\bm{\eta}_{\max}}2(\log 1/\epsilon+2)$ time and divide it into $(\log 1/\epsilon+2)$ slots of $\frac{2\thit}{1-\bm{\eta}_{\max}}$ each. By the Markov property of random walks, we know that the random walks in each of these slots are independent. So, we have the following result.
\begin{equation}
\proby{D\geq \frac{\thit}{1-\bm{\eta}_{\max}}2(\log 1/\epsilon+2)} \leq\frac{\epsilon}{4}.
\label{eq:D_bound}
\end{equation}
So, now there are two delays associated with any data packet: one is the self-delay of  $1-\bm{\eta}_{\max}$ and the second one is due to the presence of other data packets in the queue. So, the number of different delay sequences of length $s$ is at most $N^{(*)}\cdot (N^{(*)})^s \cdot \binom{D+s}{s}\cdot \binom{s+K}{s+1}$. This is because there are at most $\binom{D+s}{s}$ possibilities of choosing the intermediate path lengths $\ell_i$ such that $\sum_{i=1i}^s \ell_i \leq D $ as, despite of self-delay the number of steps is still upper bounded by $D$, and then there are $N^{(*)}$ possibilities to choose packet $p_1$. Similarly, for all other $s$ delay packets there are $N^{(*)}$ possibilities. The last factor comes from choosing a set of ranks from range $K$. Moreover, probability of choosing a delay sequence such that the ranks are distinct is $1/K^{s+1}$. So, 
\begin{align}
&\proby{\text{All $N^{(*)}$ packets get sunk in at least $D+s$ steps}}\nonumber\\
&\leq \proby{\text{There exists a Delay sequence of length $s$}}\nonumber\\
&\leq N^{(*)}\cdot (N^{(*)})^s \cdot \binom{D+s}{s}\cdot \binom{s+K}{s+1}\cdot\frac{1}{K^{s+1}}\nonumber\\
&\leq 2^{D+2s+K}\left(\frac{N^{(*)}}{K}\right)^{s+1}
\label{eq:delay_d+s}
\end{align}
If we set $K\geq 8N^{(*)}$ and $s=D+8N^{(*)}+\log 1/\epsilon-1$, then
\begin{equation}
\proby{\text{All $N^{(*)}$ packets get sunk in at least $D+s$ steps}}\leq\frac{\epsilon}{4}.
\label{eq:d+s_final}
\end{equation}
Finally, combining Eq.~\eqref{eq:D_bound} and Eq.~\eqref{eq:d+s_final} we get the desired result. 
\end{proof}

Now, since both the chains $Q^1_t$ and $Q^2_t$ operate in parallel, the expected time for the two chains to couple i.e., all red and blue packets get sunk is the maximum of the time taken by each to get their respective packets sunk. So, using Lemma~\ref{lem:old_pkt_sinking_time} for both the chains we have the expected time for $Q_t^1$, $Q_t^2$ to couple, let it be $\tau_{couple}^{1,2}=\max\{T_{N^{(red)}},T_{N^{(blue)}}\}$ as
\begin{equation}
\proby{\tau_{couple}^{1,2}\geq \frac{4\thit}{1-\bm{\eta}_{\max}} \left(\log \frac{1}{\epsilon}+2\right)+8\max\{N^{(red)},N^{(blue)}\}+\log \frac{1}{\epsilon}-1}
\leq \epsilon.
\label{eq:max_time}
\end{equation}
Note that this expected coupling time result is similar to the delay result of Leighton et al. \cite{Leighton-Combinatorica:1994}\cite{Leighton-JA:1994} depicting the pipelining behaviour of Data Collection process.

Now, to bound the distance between the two chains $Q_t^1$ and $Q_t^2$ we use the following result from Levin et al. \cite{Levin-BOOK:2009}.
\begin{lemma}[Theorem 5.2, Levin et al. \cite{Levin-BOOK:2009}]\
\label{lem:levin_coup}
Let $\{(X_t,Y_t)\}$ be a coupling with initial states $\bm{x},\bm{y} \in \mathcal{X}$ such that $X_0=\bm{x}$ and $Y_0=\bm{y}$ and coupling time defined as $\tau_{couple}:=\min \{t:X_s=Y_s \text{ for all } s\geq t\}$, then,
$$||\mathcal{P}^t[\bm{x},\cdot]-\mathcal{P}^t[\bm{y},\cdot]||_{TV}\leq P_{\bm{x},\bm{y}}\{\tau_{couple}>t\}. $$
\end{lemma}
Let $\bm{x},\bm{y} \in (\nn0)^{|V|-1}$ be the initial states of $Q_t^1$ and $Q_t^2$ chain then using Lemma \ref{lem:levin_coup} and the expected coupling time from Eq. \eqref{eq:max_time} for $||\probmt{\bm{x},\cdot}-\probmt{\bm{y},\cdot}||_{TV}\leq \epsilon$, we have 
\begin{equation}
||\probmt{\bm{x},\cdot}-\probmt{\bm{y},\cdot}||_{TV}\leq 2^{\left(\frac{\frac{8\thit}{1-\bm{\eta}_{\max}}+8\max\{N^{(red)},N^{(blue)}\}-1}{\frac{4\thit}{1-\bm{\eta}_{\max}}+1}\right)}\cdot \left(\frac{1}{2}\right)^{\frac{t}{\frac{4\thit}{1-\bm{\eta}_{\max}}+1}}
\label{eq:geo_rate_q1_q2}
\end{equation}
Now, assume the stable data rate at which we are running these stochastic processes is $\beta=(1-\delta)\beta^*$ where $\beta^*$ is the critical data rate and $\delta \in (0,1)$. Also, from linearity of $\bm{\eta}$ (see Eq.~\eqref{eq:vector_queue_eq_1}) we know $\bm{\eta}_{\max}^\beta=\frac{\beta}{\beta^*}$ as $\bm{\eta}_{\max}^{\beta^*}=1$ and hence, we have $1-\bm{\eta}_{\max}^\beta=\delta$. Using this in Eq.~\eqref{eq:geo_rate_q1_q2} we get the desired result.
\end{proof}
To use Theorem~\ref{thm:geo_ergod_gen} to prove the geometric ergodicity result (Corollary~\ref{cor:geo_ergod_pi}) we pick $\bm{y}$ according to the stationary distribution $\pi$ of the Data Collection process Markov chain.

\begin{proof}
[Proof of Corollary~\ref{cor:geo_ergod_pi}]
Let us consider two instances of Data Collection process $Q_t^1$ and $Q_t^2$ such that the former starts from some finite state $\bm{x}\in (\nn0)^{|V|-1}$ and the latter starts from stationarity i.e., initially all queues in $Q_t^1$ are occupied by some finite number of packets and that of $Q_t^2$ are filled according to the stationary distribution $\pi$. Then, from Theorem~\ref{thm:geo_ergod_gen} we have 
\begin{equation}
||\probmt{\bm{x},\cdot}-\pi||_{TV}\leq 2^{\left(\frac{8\thit-\delta}{4\thit+\delta}\right)}\cdot 2^{\left(\frac{8\max\{N^{(\bm{x})},N^{(\pi)}\}\delta}{4\thit+\delta}\right)}\cdot \left(\frac{1}{2}\right)^{\frac{\delta}{4\thit+\delta}\cdot t }
\label{eq:geo_rate_0_pi}
\end{equation}
where $\thit$ is the worst-case hitting time of random walk on graph, $N^{(\bm{x})}$ and $N^{(\pi)}$ are the total number of data packets in state $\bm{x}$ and at stationarity respectively and $\delta$ is the relative distance from the critical data rate. Now, if we compare Eq.~\eqref{eq:geo_rate_0_pi} with the Definition~\ref{def:geo_ergod}  (Eq.~\eqref{eq:geo_ergod_cond}) we prove geometric ergodicity property for the Markov chain $Q_t^\beta$.

Now for random variable $N^{(\pi)}$, let $\expt{N^{(\pi)}}$ be its expectation i.e., the expected number of data packets in $Q_t^\beta$ at stationarity which by Little's law \cite{Little-OR:1961*} is equal to the product of the data generation rate and the expected latency of a data packet to reach the sink at the stationarity i.e., $\expt{N^{(\pi)}}=\frac{\beta\thit}{1-\bm{\eta}_{\max}^\beta}=\frac{(1-\delta)\beta^*\thit}{\delta}$ (from linearity of $\bm{\eta}$ and $\beta=(1-\delta)\beta^*$) where $\beta^*$ is the critical data rate and $\delta \in (0,1)$. Now, let $\epsilon_{\bm{x}} = \max \left\{\alpha \in [0,1] : N^{(\bm{x})} \leq \frac{(1-\delta)\beta^*\thit}{\delta} \cdot \left(\log \frac{1}{\alpha} + 1\right) \right\}$. So by the definition of $\epsilon_{\bm{x}}$ we have two regimes: $\epsilon\leq\epsilon_{\bm{x}}$ where the $\expt{N^{(\pi)}}\cdot \left(\log \frac{1}{\alpha} + 1\right)$ term is dominant and $\epsilon>\epsilon_{\bm{x}}$ where the $N^{(\bm{x})}$ is dominant.

For the simple case of $\epsilon>\epsilon_{\bm{x}}$, using Eq.~\eqref{eq:geo_rate_0_pi} we have
\begin{equation}
||\probmt{\bm{x},\cdot}-\pi||_{TV}\leq 4 \cdot 2^{\left(\frac{8N^{(\bm{x})}\delta}{4\thit + \delta}\right)}\cdot \left(\frac{1}{2}\right)^{\frac{\delta}{4\thit+\delta}\cdot t }.
\label{eq:pf_geo_erg_case1}
\end{equation}
Similarly for $\epsilon\leq\epsilon_{\bm{x}}$ we have 
\begin{equation}
||\probmt{\bm{x},\cdot}-\pi||_{TV}\leq 4 \cdot 2^{\left(2(1 - \delta)\beta^* (\log 1/\epsilon + 1)\right)} \cdot \left(\frac{1}{2}\right)^{\frac{\delta}{4\thit+\delta}\cdot t }.
\label{eq:pf_geo_erg_case2}
\end{equation}
Setting the RHS of Eq.~\eqref{eq:pf_geo_erg_case2} to $\epsilon$ and solving for $t$ we get that
\begin{equation}
||\probmt{x,\cdot}-\pi||_{TV}\leq 2 \cdot 2^{\left(\frac{1}{2(1 - \delta)\beta^* + 1}\right)} \cdot \left(\frac{1}{2}\right)^{\frac{\delta}{(4\thit+\delta)(2(1 - \delta)\beta^* + 1)}\cdot t}.
\label{eq:pf_geo_erg_case2_final}
\end{equation}
Combining \eqref{eq:pf_geo_erg_case1} and \eqref{eq:pf_geo_erg_case2_final} gives us the result. 

We observe that if we set $\bm{x}$ to $\bm{0}$ (all zeros), i.e., all queues are initially empty, then $\epsilon_{\bm{0}}$ is 1 so only Eq.~\eqref{eq:pf_geo_erg_case2_final} applies and we determine the mixing time by setting the RHS to $1/M$ for a given value of $M > 0$.
\end{proof}

\section{The connection to algorithms and some future directions}
\label{sec:concl}
The fact that the Data Collection Process mixes fast to its stationary distribution when started from the all-empty setting can be exploited to solve systems of equations such as Eq.~\eqref{eq:steady-state_pot} simply by allowing the process to get close enough to stationarity and then estimate the $\bm{\eta}$ by keeping track of the number of time slots for which each queue is occupied. This opens up the possibilities of distributed algorithms for effective resistance and other problems, some of which we have explored in~\cite{Gillani-arXiv:2019}. Even if we consider graph problems on very large graphs, Laplacian systems of equations become tractable via this method since random walks can be simulated very fast in modern computing systems for graphs with nodes in the millions (see, e.g.,~\cite{sengupta-icde:2019}).

The key shortcoming of our work is that the Data Collection Process in the subcritical region models only one-sink Laplacian systems of equations. A model that captures the full generality of Laplacian systems of equations will open a more general class of problems that can be attacked algorithmically using this method.

\bibliographystyle{plain}
\bibliography{main}
\end{document}